# Darkness in interlayer and charge density wave states of 2H-TaS$_2$

Luigi Camerano,[1,][*] Dario Mastrippolito,[2] Debora Pierucci,[2] Ji Dai,[3] Massimo Tallarida,[3] Luca Ottaviano,[1,4] Gianni Profeta,[1,4] and Federico Bisti[1]

[1]*Department of Physical and Chemical Sciences, University of L'Aquila, Via Vetoio, 67100 L'Aquila, Italy*
[2]*Sorbonne Université, CNRS, Institut des NanoSciences de Paris, 4 place Jussieu, 75005, Paris, France*
[3]*ALBA Synchrotron Light Source, Cerdanyola del Vallès 08290, Barcelona, Spain*
[4]*CNR-SPIN L'Aquila, Via Vetoio, 67100 L'Aquila, Italy*

The wave-like nature of electrons is evident from quantum interference effects observed during the photoemission process. When there are different nuclei in the unit cell of a crystal and/or structural distortions, photo-electron wavefunctions can interfere, giving rise to peculiar intensity modulation of the spectrum, which can also hide energy states in a photoemission experiment. The 2H phase of transition metal dichalcogenides, with two nonequivalent layers per unit cell and charge density wave distortion, is an optimal platform for such effects to be observed. Here, we discover undetectable states in 2H-TaS$_2$, interpreting high-resolution angular resolved photoemission spectroscopy considering interference effects of the correlated electron wave functions. In addition, phase mismatching induced by the charge density wave distortion, results in evident signature of the phase transition in the photoemission spectrum. Our results highlight the importance of quantum interference, electronic correlations and structural distortion to understand the physics of layered materials.

Undetectable states by spectroscopy techniques are called dark states [1–4]. In a condensed matter system the emergence of dark states is driven by quantum interference between sublattice degrees of freedom [1]. A striking evidence of quantum interference is found in graphene, where interference between the A and B sublattices prevents access to the both linear Dirac bands using a specific linearly polarized light [5–7]. Other remarkable examples are materials hosting charge density waves (CDWs) such as Kagome lattices [8–12], topological metals [9, 13, 14] and transition metal dichalcogenides (TMDs) [15–18]. Indeed, CDWs break the translational symmetry of the lattice and induce a new periodicity in real space, resulting in a folding of the original electronic band structure in the reciprocal space. In these materials, structural distortions introduce slight phase mismatching, leading to a suppression of the intensity of the folded bands in the angular resolved photoemission (ARPES) spectra. Among 2D TMDs, 2H-TaS$_2$ is particularly suitable for such effects to be evident. The 2H phase of TMD is characterized by a glide-mirror symmetry between the layers (translation of half lattice constant followed by a reflection) resulting two pairs of chalcogenide sublattices, which can produce destructive quantum interference. In addition, 2H-TaS$_2$ hosts an incommensurate CDW transition below T$_{CDW} \simeq$ 75 K [19–22] also making the study of CDW driven quantum interference possible in photoemission experiments. The observation and description of the electronic structure of TMD is a fundamental step to the comprehension of their electronic, structural, magnetic and superconducting phases in order to integrate them in the present technology. To this scope, very recently TaS$_2$ received special attention because it hosts multiple ground states, depending on its structural phase. Its most stable phase at ambient condition (2H-TaS$_2$) [23] is metallic, and at low temperature the CDW phase compete with a superconductive one [24]. The CDW should originate from the strong electron-phonon coupling, since the nesting vectors of neither of the two Fermi surfaces at the Γ point account for the CDW wave periodicity [20, 21, 25]. A remarkable increase of the superconducting [26–28] and CDW critical temperatures [29] have been observed going from the bulk to monolayer phase, highlighting intriguing dependencies of the electronic properties on the dimensionality of the system. On the contrary, the low energy 1T phase, characterized by the trigonal stacking, develops a CDW transition towards a "star of David" distortion eventually leading to a Mott insulating phase [30–32]. Finally, an alternate stacking between 1H-TaS$_2$ and 1T-TaS$_2$ monolayers, namely the 4H$_b$-TaS$_2$ phase, was found to host intricate interplay between correlated Mott and Kondo physics, topology and chiral superconductivity [33–39] whose origins are still debated from a theoretical and experimental point of view.

The understanding of the correlation effects in the electronic properties of TaS$_2$ is thus crucial to describe this complex phase diagram, which is further enriched by the dependence on the sample dimensionality, stacking, and quantum interference effects.

In this study we unveil the electronic bulk structure of 2H-TaS$_2$ in the full three-dimensional (3D) Brillouin Zone (BZ) by ARPES measurements carefully combined with the state-of-the-art first-principle calculations. We provide a comprehensive interpretation of the quantum interference effects driven by the sublattices and CDW through an unfolding technique based on the combination of the complex orbital projected Kohn-Sham wave functions, revealing the presence of dark states dispers-ing along the $k_z$ direction and loss of intensity in the ARPES spectrum caused by wavefunction phase mismatching present in the CDW phase. Bulk 2H-TaS$_2$ is formed by stacked 1H-TaS$_2$ monolayers (MLs) in an ABA sequence connected by a glide-mirror symmetry (Fig. 1). The 1H-TaS$_2$ ML consists of a hexagonal lattice of Ta atoms in trigonal prismatic coordination with the chalcogenides [40] (see Fig. 1a left and right panel). Its band structure calculated within the Density Functional Theory (DFT) (see Methods in Supplementary Material (SM)), Fig. 1b, is characterized by metallic Ta-$d$ states crossing the Fermi level as shown in Fig. 1b. Just below these Ta-$d$ states, sulfur-$p_z$ states are found, while the in-plane S-$p_{x,y}$ are lower in energy.

[*] email: luigi_camerano@outlook.it

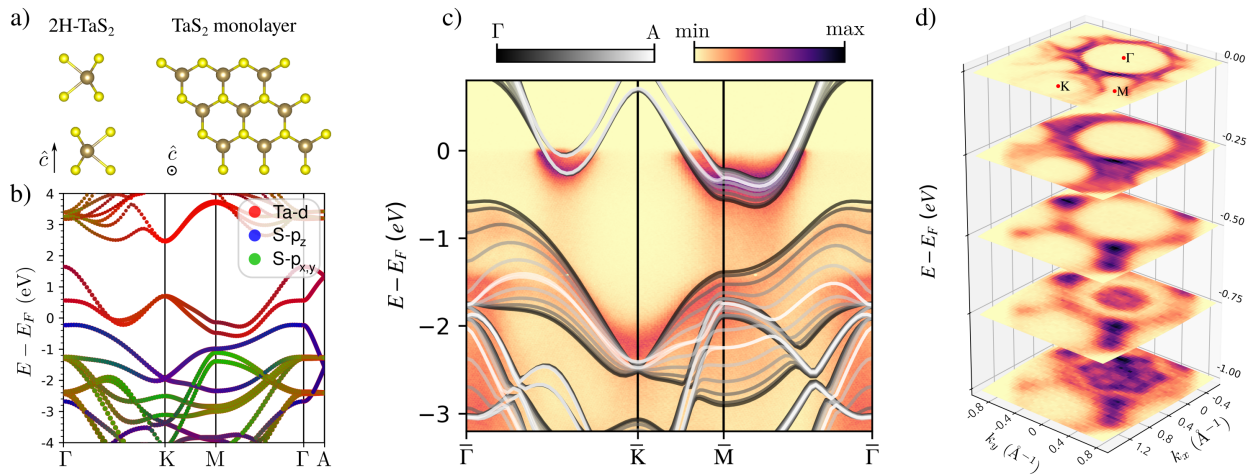

FIG. 1. a) Left panel: Primitive cell of 2H-TaS$_2$ showing the glide-mirror symmetry along the c axis. Right panel: top view of the crystal structure of 1H-TaS$_2$ monolayer: hexagonal lattice of Ta atoms (brown spheres) in edge-sharing trigonal coordination with the chalcogenide S atoms (yellow spheres). b) 2H-TaS$_2$ bands structure calculated with PBE+SOC, projected on Ta-$d$, S-$p_z$, and S-$p_{x,y}$ orbitals. c) HSE band structure superimposed to the high-resolution map of the electronic band structure of 2H-TaS2 crystals collected at h$\nu$ = 75 eV and T = 50 K along the $\bar{\Gamma}\bar{K}\bar{M}\bar{\Gamma}$ direction where the Fermi level is located at the zero of the binding energy. The grey colormap represents different $k_z$ cuts of the band structure (from $\Gamma$ to $A$). d) Isoenergy contours along the $\bar{\Gamma}\bar{K}\bar{M}$ plane.

In order to validate our theoretical description, we performed high-resolution ARPES spectra of the bulk 2H-TaS$_2$ band structure along the $\bar{\Gamma}\bar{K}\bar{M}\bar{\Gamma}$ lines of the surface BZ, using a photon energy of 75 eV and at temperature of $T = 50$ K. The comparison with first-principle HSE calculations [41, 42] (see methods in SM for further details) confirms the theoretical band structure. We note that Ta-$d$ band dispersion around the M point and the position of the S-$p_z$ state are dependent on the exchange-correlation functional used in the calculation. Specifically for the S-$p_z$ state, while local functionals like GGA [43] underestimates its binding energy (see Fig. S1 in SM) the HSE functional predict it in agreement with the experimental measurement. The effect of the exchange-correlation has remained unexplored in Ref. [19, 44] because the ARPES spectra were obtained in the ML case, where the S-$p_z$ band is found at lower binding energy (see Fig. S3 and S4 in SM) than the bulk, and have been not compared with the calculated band structure. The importance of the proper description of the exchange-correlation effects is not only limited to 2H-TaS$_2$ but needed also in other S-based TMDs [45] and it has been often overlooked [46–48].

The broadening of the spectra reflects the strong dispersion along the out-of-plane momentum, coming from the layer hybridization. This effect is precisely reproduced by our calculations, when we project the states along the $k_z$ direction (greyscale colored lines in Fig. 1c). In Fig. 1d we report the isoenergy map along the $\bar{\Gamma}\bar{K}\bar{M}$ plane highlighting the presence of a hole pocket at $\bar{K}$. The topology of the Fermi surface changes only at $E - E_F =$ -0.75 eV where the S-$p_z$ band gives rise to an additional $\Gamma$-centered sheet, while the cut at $E - E_F =$ -1 eV reveals a more complex structure due to the dispersion of the S-$p_z$ band. Therefore, we pointed out a sensible stacking and correlation effects on the electronic band structure of TaS$_2$: ($i$) a slight modification of Ta-$d$ states near the Fermi level and ($ii$) an overall rigid shift of the filled S-$p$ bands of about 0.4 eV due to non-local exchange-correlation effects (see SM for further details).

However, a careful analysis of the experimental ARPES spectra in the full Brillouin zone, reveals unexpected features related to the sublattice degree of freedom of the 2H phase. In the 2H phases of TMDs, the chalcogenide atoms form two pairs of sublattices, each one belonging to a different layers. Accessing the $k_z$ dispersion of the bands by varying the photon energy from 55 to 95 eV, see Fig. 2, we will demonstrate that this symmetry is the source of quantum interference effects possibly giving rise to dark states. The isoenergy contour (at $E - E_F =$ -1 eV) in the $\Gamma AK$ plane is shown in Fig. 2a showing also the $k_z$-path along $\Gamma - A$ line, whose band dispersion is reported in Fig. 2b with $p$-polarized light and in Fig. 2c with $s$-polarized light. Starting from $\Gamma$ and going to $A$ point we recognize a down-dispersing parabola which is then not replicated in the first following $\Gamma$ point, but it reappears only on the second one (Fig. 2b). Such apparent doubling of the BZ size is due to an interference effect (see below) stemming from the presence of a glide-mirror symmetry operator in the unit cell of 2H-TaS$_2$ with affects the dipole matrix elements. Analog photoemission interference effects were already observed for graphene [6, 49, 50], iron pnictides [51], WTe$_2$ [52] and CrO$_2$ [53]. However, it is widely recognized that changing the light polarization (and thus the dipole matrix element) makes detectable the previously unobserved signals. This is not the case of 2H-TaS$_2$ as is shown in Figs. 2b and c.

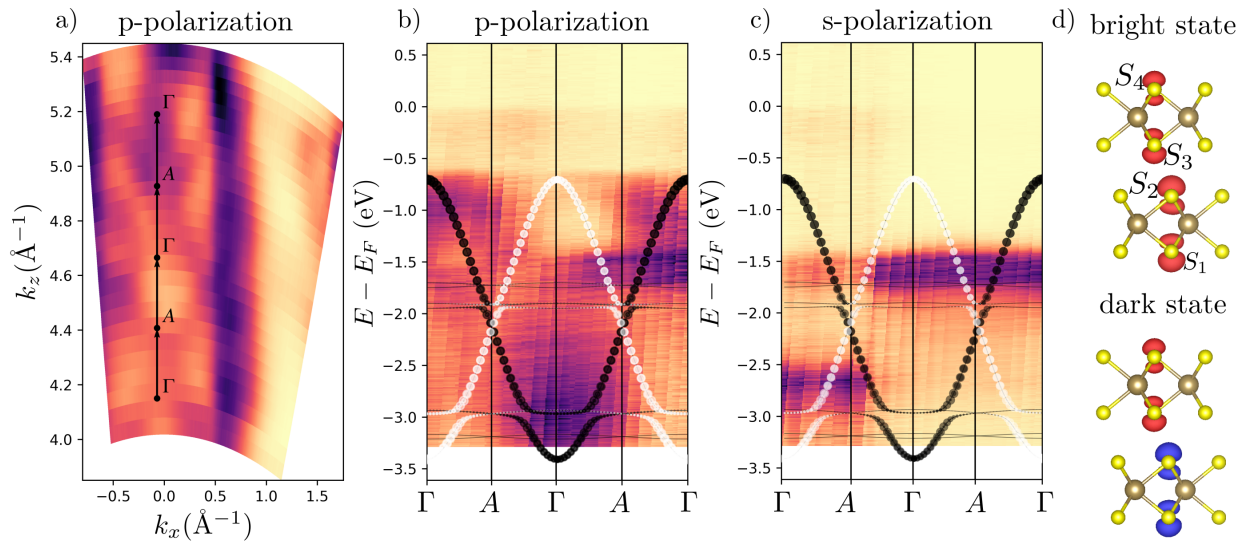

FIG. 2. Photon energy scan (from 55 to 95 eV). a) Isoenergy contour along the $\Gamma AK$ plane acquired with $p$-polarized light. b) Band dispersion along the $\Gamma A$ direction. The unfolded band structure is represented with the point size proportional to the total spectral weight of the projection on S-$p_z$ orbital (in black the "even" combination, in white the "odd" one; see the main text) acquired with $p$-polarized light. c) Band dispersion along the $\Gamma A$ direction acquired with $s$-polarized light. d) A schematic of the bright and dark state. In the case of the bright state, all the S atoms have the same phase factor, while in the case of the dark state $S_1$ and $S_2$ have a $\pi$-phase.

In order to correctly understand the ARPES spectrum, we calculate the photoemission matrix element $M$ introducing an unfolding technique based on the interference of the complex orbital-projected Kohn-Sham wave function. Indeed, adapting the expression of $M = \langle \psi_f | \mathbf{A} \cdot \mathbf{p} | \psi_i \rangle$ from Refs. [1, 54] for our case into a DFT framework and simplifying the initial state as the projected wavefunction on the $p_z$ atom-centered orbitals we can write:

$$M \sim \langle \psi_f | (\mathbf{A} \cdot \mathbf{p}) \sum_j C_j^{n\mathbf{k}} | p_{z,j}^{S_j} \rangle \qquad (1)$$

Supposing $|\psi_f\rangle$ an even function, from Eq. (1) it is clear that $M$ is determined by the electron momentum $\mathbf{p}$, whose direction is selected by the light polarization contained in $\mathbf{A}$, acting on the linear combination of the $C_j^{n\mathbf{k}}$ coefficients ($C_j^{n\mathbf{k}} = \langle p_z^{S_j} | \psi_{\mathbf{k}n}^{KS} \rangle$, where $j$ is the sublattice index, see SM for further details). From Eq. (1) quantum interference arises from the summation of $C_j^{n\mathbf{k}}$ coefficients that are determined by the symmetry of the crystal.

In the case of 2H-TaS$_2$, the $k_z$ dispersing bands shown in Fig. 2b-c originates from S-$|p_z\rangle$ orbitals (see Fig. 2d for the corresponding first-principle charge density map). Weighting the band structure with $|M|^2$, as shown in Fig. 2d, we correctly recover the periodicity of the ARPES spectra as measured with $p$-polarized light (see Fig. 2b where the "even" combination is a bright state reported with black dots). On the contrary, with $s$-polarized light, the $p_z$-like bands cannot be detected, thus making the 'odd' combination ($S_1$ and $S_2$ out of phase with $S_3$ and $S_4$) an interlayer dark state (reported in white in Figs. 2b-c, see SM for further details) in analogy with the definition coined in Ref. [1]. However, it is important to underline that our case differs from that one present Ref. [1] because it refers to a out-of-plane dispersing band.

As mentioned before, interference from sublattices is not the only source of modulation of the photoemission intensity: structural distortions, which increase the periodicity of the unit cell, can suppress the spectral weight on the folded bands ($|M|^2$). For example, CDW distortion induces phase mismatching in the Kohn-Sham wavefunctions strongly reducing $M$ term at specific points in the BZ, eventually resulting in gaps observed in the unfolded bands, which we refer to as pseudo-gaps. In Fig. 3a we report the same Fermi surface map as reported in Fig. 1d but highlighting the relevant intensity variation of the photoemission signal along the $\bar{\Gamma}\bar{K}$ (indexed as zone 1) and $\bar{K}\bar{M}\bar{K}$ directions (indexed as zones 2 and 3) around the $\bar{K}$ hole pocket. The band dispersions in these regions are reported in Figs. 3b-c (left panels), showing a clear suppression of the intensity close to the Fermi energy, as indicated by the black circles in the figures. The calculated band structure of the undistorted 2H-TaS$_2$ (see Fig. 1c) does not account for these features, because they are ascribed to the coupling of the electrons with the CDW distortion. In order to verify this hypothesis, we perform first-principle calculations of 2H-TaS$_2$ in the CDW phase using a $3 \times 3 \times 1$ supercell [55], close to observed incommensurate wave-vector (see DFT methods for further details, the relaxed atomic positions are shown in Fig. S4). Then, using Eq. 1 we unfold the supercell band structure into the $1 \times 1$ BZ at $k_z = 0$. The results are shown in Fig. 3b-c (right panel), in which we also highlight with the vertical grey arrow the effect of out-of-plane dispersion, whose effects at the $\Gamma$ points were presented in Fig. 2 and discussed in terms of the $k_z$ dispersion of the band.

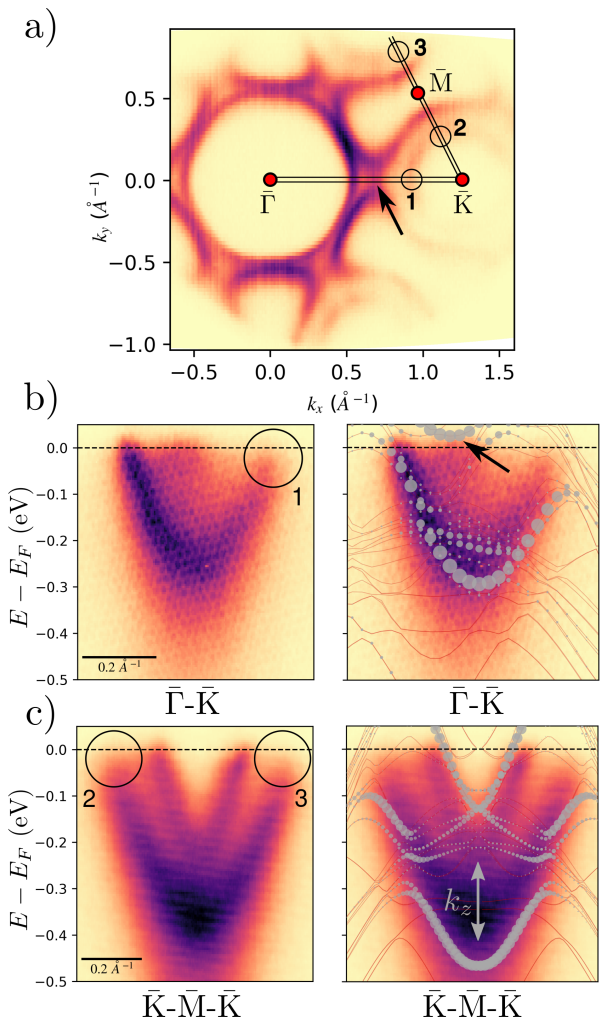

FIG. 3. a) Fermi surface map collected at $h\nu=$ 75 eV b)-c) Left panels: High-resolution detail of the electronic band structure along the $\bar{\Gamma}\bar{K}$ and $\bar{K}\bar{M}\bar{K}$ directions, respectively. The circles are in correspondence with the highlighted spots in panel a). Right panels: comparison between ARPES spectra and PBE band structure in the CDW phase unfolded on the $1 \times 1$ cell. The unfolded band structure is represented with the point size (in grey) proportional to the total spectral weight of the projection on Ta-$d$ orbitals. The black arrow in panel b) shows a band shifting in correspondence of the black arrow in panel a), while the grey arrow in panel c) shows the broadening due to the $k_z$ dispersion.

The unfolded bands calculated in the distorted phase nicely account for the loss of spectral weight observed in the experiment: pseudo gaps are opened where the experimental band show reduced intensity (the regions indicated in Fig. 3a as 1, 2, and 3). Focusing on right panels of Figs. 3b-c, we can clearly see how the calculated band dispersion modifies with respect to the undistorted bands (see Fig. 1b). We can also estimate the magnitude of the pseudo-gap from the experimental energy dispersive curves finding $\Delta_{\bar{\Gamma}\bar{K}} = \Delta_{\bar{K}\bar{M}\bar{K}} \sim 90$ meV within our experimental resolution.

Another effect of the CDW is the emptying of a band, as indicated by the black arrow in the right panel of Fig. 3b. Interestingly, ARPES measurements in the monolayer 1H-TaS$_2$ detect a similar effect at low temperatures in the CDW phase [19], without relating it to a CDW pseudo-gap opening. As confirmation of the origin of the pseudo-gaps, we underline that they are not present when CDW is suppressed by the interaction of TaS$_2$ with a substrate [44]. Moreover, we note that similar pseudo-gaps are present along the $\bar{\Gamma}\bar{K}$ direction in 4H$_b$-TaS$_2$ as recently reported in Ref. [56]. This indicates the coexistence of multiple CDW orders in this compound (besides the $\sqrt{13} \times \sqrt{13}$ in the 1T layer), as measured in a very recent STM experiment [57] where a $3 \times 3$ modulation is revealed by the STM map which can be ascribed to the 1H layer.

In conclusion, we have presented a detailed analysis of the TMD 2H-TaS$_2$ electronic structure by combining high-resolution ARPES and state-of-the-art DFT calculations. We have introduced a novel paradigm for interpreting $k_z$ dispersion in layered materials composed of sublattice pairs connected by symmetry operations which allow us to discover interlayer and CDW-driven dark states in 2H-TaS$_2$. These interlayer dark state were already observed in other materials such as 2H-NbSe$_2$ [45] and 2H-TaSe$_2$ [58] and are now completely understood within our formalism. Moreover, our results unveil the impact of exchange-correlation effects in its ARPES spectrum, particularly affecting the S-$p_z$ bands responsible for the interlayer coupling and the Ta-$d$ bands crossing the Fermi level. We provide useful insights into the intricate interplay between structure, electronic properties, and quantum interference effects in 2D materials, demonstrating that an accurate description of the electronic spectrum can only be achieved using first-principles calculations beyond local density approximations, including the CDW distortion and accounting for interference effects that modulates the photoemission spectra. Our proposed unfolding procedure demonstrates CDW origin from phase mismatching of the wavefunction, revealing it as a consequence of interference effects caused by lattice distortion. These results pave the way for the understanding of the complex dependencies of the physical properties on dimensionality, 1H/1T stacking, and intercalated phases in TaS$_2$ in which interlayer interaction plays an essential role.

## I. DATA AVAILABILITY STATEMENT

All data that support the findings of this study are included within the article and supplementary materials.

## II. ACKNOWLEDGEMENTS

The ARPES measurements have been done at LOREA beamline of ALBA Synchrotron Light Source. LOREA was co-funded by the European Regional Development Fund (ERDF) within the Framework of the Smart Growth Operative Programme 2014-2020. Authors thank Jordi Prat for the invaluable technical support during ARPES experiments. G.P. acknowledges support from CINECA Supercomputing Center through the ISCRA project and financial support from the Italian Ministry for Research and Education through the PRIN-2017 project "Tuning and understanding Quantum phases in 2D materials-Quantum 2D" (IT-MIUR Grant No. 2017Z8TS5B). This work was funded by the European Union-NextGenerationEU under the Italian Ministry of University and Research (MUR) National Innovation Ecosystem Grant No. ECS00000041 VITALITY-CUP E13C22001060006. F.B. acknowledges funding from the National Recovery and Resilience Plan (NRRP), Mission 4, Component 2, Investment 1.1, funded by the European Union (NextGenerationEU), for the project "TOTEM" (CUP E53D23001710006 - Call for tender No. 104 published on 2.2.2022 and Grant Assignment Decree No. 957 adopted on 30/06/2023 by the Italian Ministry of Ministry of University and Research (MUR)) and for the project "SHEEP" (CUP E53D23018380001 - Call for tender No. 1409 published on 14.9.2022 and Grant Assignment Decree No. 1381 adopted on 01/09/2023 by the Italian Ministry of Ministry of University and Research (MUR)). This work was supported by a public grant overseen by the French National Research Agency (ANR) through the grants Operatwist (No. ANR-22-CE09-0037-01), E-map (No. ANR-23-CE50-0025), and as part of the "Investissements d'Avenir" program (Labex NanoSaclay, reference: ANR-10-LABX-0035). This project has received financial support from the CNRS through the MITI interdisciplinary programs (project WITHIN).

Supplemental material and supporting information for

# Darkness in interlayer and charge density wave states of 2H-TaS2

Luigi Camerano [a], Dario Mastrippolito [c], Debora Pierucci [c], Ji Dai [d], Massimo Tallarida [d], Luca Ottaviano [a,b], Gianni Profeta[a,b] and Federico Bisti [a]

[a] Department of Physical and Chemical Sciences, University of L'Aquila, Via Vetoio 67100 L'Aquila, Italy
[b] CNR-SPIN L'Aquila, Via Vetoio 67100 L'Aquila, Italy
[c] Sorbonne Université, CNRS, Institut des NanoSciences de Paris, 4 place Jussieu, 75005, Paris, France
[d] ALBA Synchrotron Light Source, Cerdanyola del Vallès 08290, Barcelona, Spain

CONTENTS

## I. METHODS

ARPES experiments were performed at the LOREA beamline of the ALBA synchrotron (Spain) using an MBS A1 hemispherical analyzer with a horizontal slit. The photon energies used were in the 55–95 eV range with linear horizontal polarized light ($p$ polarization) and vertical polarized light ($s$ polarization), and the sample temperature was T = 50 K. The samples, commercial crystals from HQ graphene, were exfoliated in the UHV chamber using a Kapton tape.
Density functional theory calculations were performed using the Vienna ab-initio Simulation Package (VASP) [1, 2], using both the generalized gradient approximation (GGA), in the Perdew-Burke-Ernzerhof (PBE) parametrization for the exchange-correlation functional [3] and HSE06 hybrid functional [4–6]. Interactions between electrons and nuclei were described using the projector-augmented wave method. Energy thresholds for the self-consistent calculation was set to $10^{-5}$ eV and force threshold for geometry optimization $10^{-4}$ eV Å$^{-1}$. A plane-wave kinetic energy cutoff of 500 eV was employed. The Brillouin zone was sampled using an $8 \times 8 \times 4$ for the bulk and a $12 \times 12 \times 1$ Gamma-centered Monkhorst-Pack grid. To account for the on-site electron-electron correlation we used the GGA+U approach with an effective Hubbard term $U = 4$ eV as calculated with linear response theory [7]. The 2H-TaS$_2$ lattice parameter are set to the experimental ones: $a = 3.31$ Å and $c = 12.07$ Å [8]. For the monolayer, a vacuum region of 15 Å is adopted to decouple the different layers. The structural relaxation of the internal position using HSE06 functional was performed using the FHI-AIMS simulation package [9, 10], which is an accurate all-electron full-potential electronic structure package based on numeric atom-centered orbitals, with so-called "tight" computational settings. The screened hybrid functional HSE06 with the mixing factor $\alpha = 0.25$ and screening parameter $\omega = 0.11$ Bohr$^{-1}$ was used for the exchange-correlation energy.
A $3 \times 3 \times 1$ supercell was used to investigate the CDW structural transition. As it has been experimentally shown that in 2H-TaS$_2$ the CDW phase is incommensurate, we applied a perturbation on the atomic position according to the CDW displacement predicted by Ref. [11] and then we fully relaxed the atomic position. The Brillouin zone was sampled using a $5 \times 5 \times 5$ Gamma-centered Monkhorst-Pack grid with a smearing parameter of $\sigma = 0.01$ eV.

## II. CORRELATION EFFECTSIN THE BAND STRUCTURE OF 2H-TAS$_2$

In this section we analyze the role of correlations in the band structure of 2H-TaS$_2$. Indeed, by extending our calculation using non-local potentials (HSE), we observe a relevant modification of the band structure as reported in Fig. S1. HSE opens the gap between sulfur-derived states and Ta-$d$ conduction bands. We also notice a slight modification on the band crossing the Fermi level (Ta-$d$ character bands around the M point), making them more dispersing in energy in the case of HSE rather than PBE. We ascribe these effects to the limitation of local density functionals in describing the localized Ta-$d$ electrons close to the Fermi level. Indeed, the calculation of the Hubbard repulsion ($U$) using linear response theory [7, 12] (see Fig. S2) gives $U= 4.12$ eV also for the 2H-TaS$_2$. The effect on the band structure of the Hubbard correction is very similar to the HSE calculation for the filled state (see Fig. S1) while leaves almost unchanged the empty states.

The primary consequence of accounting for correlations beyond PBE is an overall shift of approximately 0.4 eV in the sulfur-derived bands. This is a general effect not only limited to 2H-TaS$_2$ but present also in other S-based TMDs [13] and it is often overlooked [14–16]. We underline that the electronic correlations in Ta-$d$ bands are the fundamental element in the 1T phase give rise to its very rich electronic landscape [17–19].

Analog effects can be traced back to the monolayer, as we note an evident modification of the dispersion of the S-$p_z$ band (see Fig. S3), while the Ta-$d$ bands below the Fermi level are only partially modified along the $\Gamma - M$ direction. Finally, we discuss the main differences between monolayer and bulk phase band structure in TaS$_2$. In Fig. S4 we report crystal structure, BZ and band structure comparison between monolayer and bulk $H$-phase of TaS$_2$.

The main difference regarding the Ta-$d$ bands below the Fermi level (i.e. visible in ARPES) between monolayer and bulk are around the M point. As discussed in the main text and visible in Fig. 1c and Fig. 3c this discrepancy is rationalized by noting an emergence of $k_z$ dispersion mainly due to the interlayer hybridization. Here, correlation effects modify the band structure (see Fig. S1), signaling the importance of the proper inclusion exchange-correlation effects in the description of interlayer hybridization. Similar effects for the same band but above the Fermi level can be seen at the $\Gamma$ point, where a dispersion along the $\Gamma$A appears. However, the most relevant failure of generalized gradient approximation (GGA) regards the $S$-p$_z$ bands (see Fig. S1) that give rise to an out-of-plane dispersive band completely absent in the case of monolayer (see Fig. S4). Indeed, GGA functionals completely miss the energy of this state (see Fig. S1).

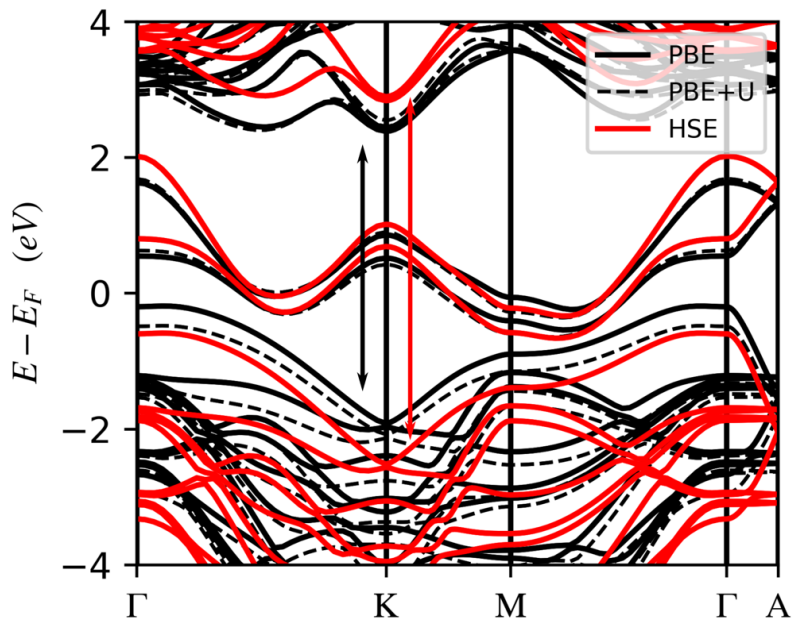

FIG. S1. PBE+SOC, PBE+$U$+SOC, HSE+SOC bulk band structure.

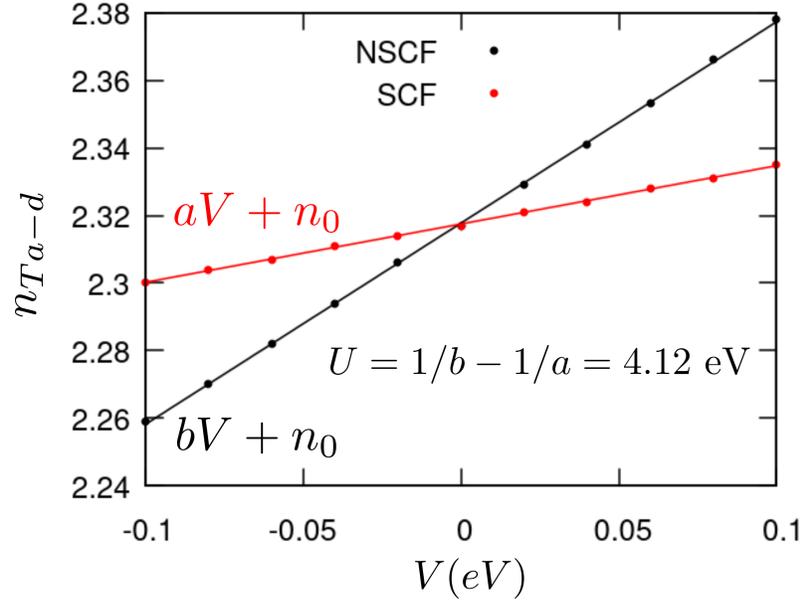

FIG. S2. Self-consistent (SCF, in red) and non-self-consistend (NSCF, in black) response of the charge occupation $n_{Ta-d}$ to an external potential V. The black and red lines are the results of the fitting procedure.

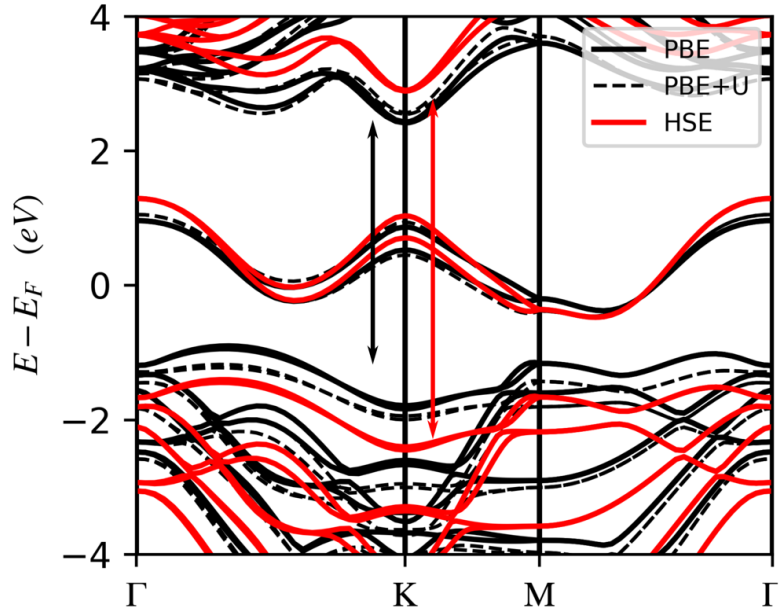

FIG. S3. PBE+SOC, PBE+$U$+SOC, HSE+SOC monolayer band structure.

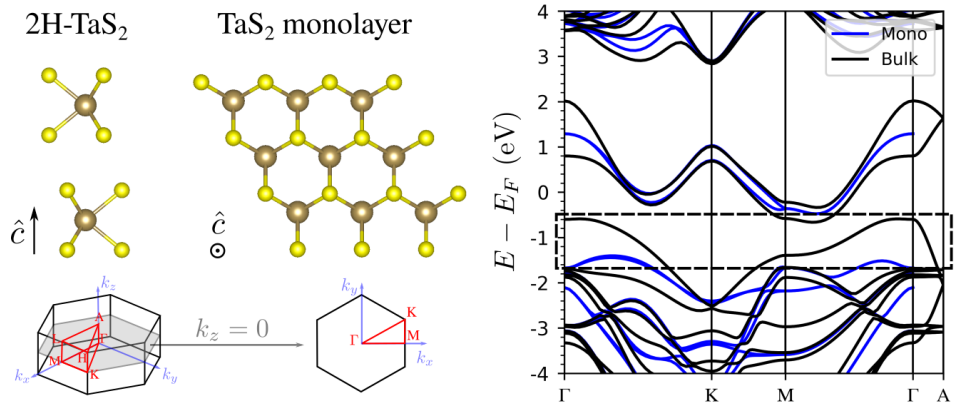

FIG. S4. Left panel: crystal structure for bulk and monolayer TaS$_2$ and corresponding BZ. Right panel: band structure comparison of monolayer and bulk phases with HSE+SOC functional.

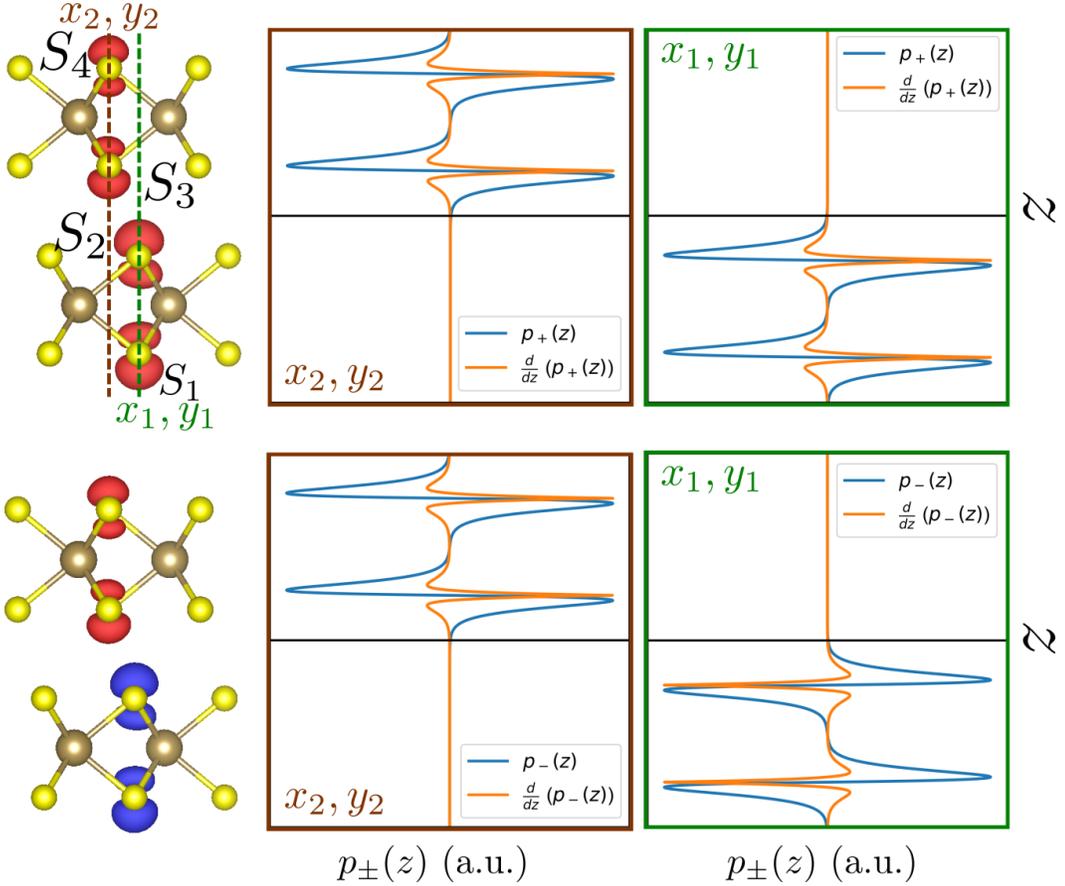

FIG. S5. Left panel: A schematic of the bright and dark state. In the case of the bright state ($p_+(z)$), all the S atoms have the same phase factor, while in the case of the dark state $S_1$ and $S_2$ have a $\pi$-phase ($p_-(z)$). Right panel: radial part of S-$p_z$ orbitals at fixed $x_1, y_1$ and $x_2, y_2$ showing the effect of the derivative on the considered atomic basis.

### III. SYMMETRY ANALYSIS FOR THE EMERGENCE OF THE DARK STATE

Let's analyze the symmetries of 2H-phase of TMD, whose TaS$_2$ is a paradigmatic example. In this material, interlayer interactions (see also Point 4), are responsible for the emergence of the out-of-plane dispersion, mainly originated by S-$p_z$-like bands. Due to the glide mirror symmetry, some eigenvalues of the Hamiltonian can be labeled as 'even' or 'odd' considering the effect of this last symmetry operation on their wavefunctions ($\psi_i$). This property strictly reflect the possibility for the state to be detecatable or undetectable in photoemission.

To see this we must look at the photemission matrix element $M = \langle \psi_f | \mathbf{A} \cdot \mathbf{p} | \psi_i \rangle$.

Considering the sample surface orientation with respect to the incoming photons in our experiment ($z$-axis $\parallel c$ direction), we can write:

$$\mathbf{A} = \lambda_x \hat{\mathbf{x}} + \lambda_y \hat{\mathbf{y}} + \lambda_x \tan(\gamma) \hat{\mathbf{z}} \qquad (1)$$

where $\gamma$ is the incident angle and the light polarization is determined by the vector $\lambda = (\lambda_x, \lambda_y, \lambda_x \tan(\gamma))$. The $p$-polarized light is the one with $\lambda_x \neq 0$ and $\lambda_y = 0$, whether $s$-polarized light is with $\lambda_x = 0$ and $\lambda_y \neq 0$. In order to demonstrate that the 'odd' state is an interlayer dark state we write $M$ considering the momentum operator in real space:

$$M \sim \int\int\int \psi_f^* (A_x \partial_x + A_y \partial_y + A_z \partial_z) \psi_i \, dx dy dz \qquad (2)$$

In the case of the interlayer states, the initial wavefunction ($\psi_i$) can be written as a linear combination of the $p_z$ orbital centered in the sulfur atom at the $j$ position ($p_z^{S_j}$). The 'even' combination with respect to the glide mirror symmetry is $p_+ \sim p_z^{S_1} + p_z^{S_2} + p_z^{S_3} + p_z^{S_4}$ and the 'odd' one is $p_- \sim p_z^{S_1} + p_z^{S_2} - (p_z^{S_3} + p_z^{S_4})$ (see Fig. S5). Considering $p$-polarized light (i.e. both $A_x$ and $A_z$ contribution) only the $p_+$ gives a contribution different from zero, due to odd symmetry

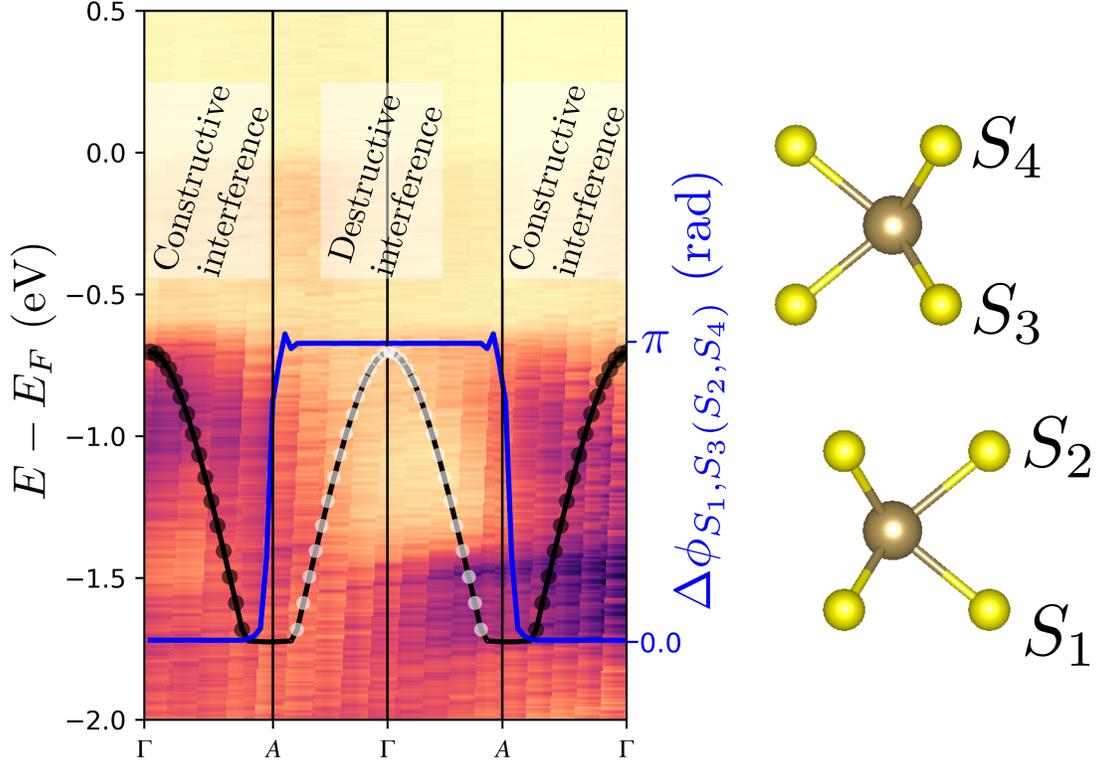

FIG. S6. Accessing the symmetry of the wavefunction by first-principle calculation. The unfolded band structure is represented with the point size proportional to the total spectral weight of the projection on S-$p_z$ orbital (in black the "even" combination, in white the "odd" one; see the main text) acquired with p-polarized light

of the derivatives (see Fig. S5) and the fact that $S_1, S_2$ are related with a mirror to $S_3, S_4$, respectively. Instead, for $s$-polarized light both $p_+$ and $p_-$ have vanishing signal due to the even symmetry with respect of $y$ component of $p_z^{S_j}$ orbitals:

$$\int\int\int \psi_f^*(A_y\partial_y)p_z^{S_j} = 0 \quad \forall j \, . \tag{3}$$

Thus, the 'odd' combination $p_-$ is an interlayer dark state, undetable with both polarizations.

From a practical point of view, one can access the symmetry of the initial wavefunction, from a DFT framework, by looking at the projection of the Kohn-Sham wavefunction $|\psi^{KS}\rangle$ on atomic basis function centered on each S atom ($\langle p_z^{S_j}|, j=1,4$). For example for $S_1, S_3$

$$\langle p_z^{S_1}|\psi^{KS}_{\mathbf{k}n}\rangle = C_1^{n\mathbf{k}} = \rho_{n\mathbf{k}}e^{i\phi_{1,n\mathbf{k}}} \, , \tag{4}$$

$$\langle p_z^{S_3}|\psi^{KS}_{\mathbf{k}n}\rangle = C_3^{n\mathbf{k}} = \rho_{n\mathbf{k}}e^{i\phi_{3,n\mathbf{k}}} \, . \tag{5}$$

Introducing the phase difference:

$$\Delta\phi_{n\mathbf{k}} = \phi_{3,n\mathbf{k}} - \phi_{1,n\mathbf{k}} \tag{6}$$

it is possible to see whether a state is 'even' ($\Delta\phi_{n\mathbf{k}} = 0$) or 'odd' ($\Delta\phi_{n\mathbf{k}} = \pi$).

At the same time, the coefficients $C_j^{n\mathbf{k}}$, where $j$ is the site index, determine if the state is bright or dark. Indeed, adapting the expression of $M$ from Refs. [20, 21] for our case and reformulating it into a DFT framework simplifying the initial state as the projected wavefunction on the $p_z$ atom-centered orbitals we can write:

$$M \sim \langle \psi_f | (\mathbf{A} \cdot \mathbf{p}) \sum_j C_j^{n\mathbf{k}} | p_{z,j}^{S_j} \rangle \qquad (7)$$

Supposing $|\psi_f\rangle$ an even function, from Eq. (7) it is clear that $M$ is determined by the momentum $\mathbf{p}$, whose direction is selected by the light polarization contained in $\mathbf{A}$, acting on the summation of the $C_j^{n\mathbf{k}}$ coefficients.
To summarize, the dark state originates from the interlayer hybridization of $S$-$p_z$ bands and cannot be detected with $p$-polarized light due to interference effects contained in the summation of Eq. 7. With $s$-polarized light, the even symmetry with respect of $y$ component of $p_z^{S_j}$ orbitals.

## IV. PSEUDO-GAP OPENING AND CDW DISTORTIONS

In Fig. S7 we estimate the pseudo-gap opening along the $\bar{\Gamma}\bar{K}$ and $\bar{K}\bar{M}\bar{K}$ directions from the energy dispersive curves (EDCs). In particular, by integrating the signal in the red and green regions in the right panel of the Fig. S4 we determine $\Delta_{\bar{\Gamma}\bar{K}} \sim \Delta_{\bar{K}\bar{M}\bar{K}} \sim 90$ meV.

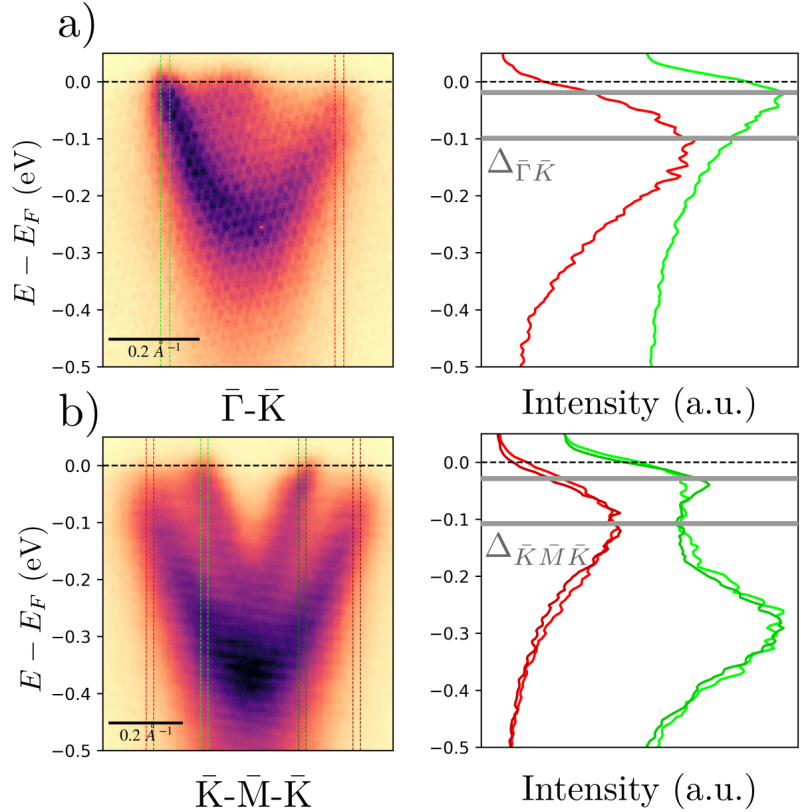

FIG. S7. a)-b) Left panels: High-resolution detail of the electronic band structure along the $\bar{\Gamma}\bar{K}$ and $\bar{K}\bar{M}\bar{K}$ directions, respectively. Right panels: corresponding EDC along the green and red cuts. The thick grey lines highlight the CDW pseudo gap.

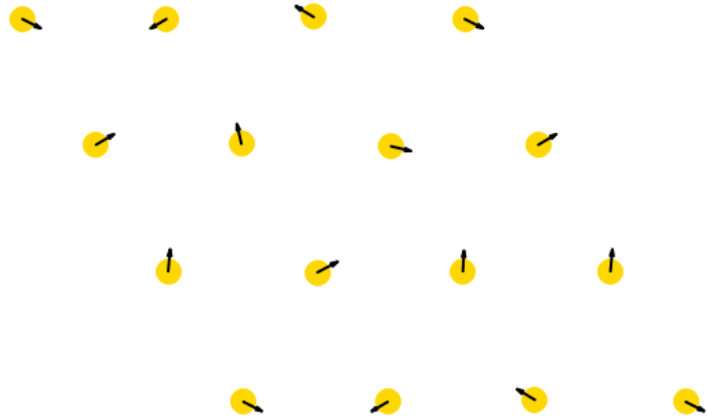

FIG. S8. Atomic displacement of the Ta atoms (yellow spheres). The black arrows represent the displacement magnified by a factor 20 to better visualize the displacement